\documentclass[aps,prl,reprint]{revtex4-2}
\usepackage[utf8]{inputenc}
\usepackage[tbtags]{amsmath}
\usepackage{amssymb}
\usepackage{graphicx}
\usepackage{mathtools}
\usepackage{dsfont}
\usepackage{enumitem}
\usepackage{bbm}
\usepackage{hyperref}
\usepackage{cases}
\usepackage[usenames,dvipsnames]{color}
\usepackage[sort&compress]{natbib}
\newcommand{\bZ}{{\textbf {Z}}}
\newcommand{\bX}{{\textbf {X}}}
\newcommand{\bx}{{\textbf {x}}}
\newcommand{\bz}{{\textbf {z}}}
\newcommand{\bbe}{{\textbf {e}}}
\newcommand{\bbf}{{\textbf {f}}}

\newcommand{\bO}{{\textbf {O}}}
\newcommand{\norm}[1]{\left\lVert#1\right\rVert} 
\newcommand{\bbR}{\mathbb{R}}

\DeclareMathOperator*{\argmin}{\mathrm{argmin}}
\DeclareMathOperator*{\extr}{\mathrm{extr}}

\begin{document}
\title{Large deviations of extreme eigenvalues of generalized sample covariance matrices}
\author{Antoine Maillard}
\email{antoine.maillard@ens.fr}
\affiliation{Laboratoire de Physique de l'\'Ecole Normale Sup\'erieure, ENS, Universit\'e PSL, Paris, France}

\begin{abstract}
  We present an analytical technique to compute the probability of rare events in which the largest eigenvalue of a random matrix is
atypically large (i.e.\ the right tail of its large deviations). The results also transfer to the left tail of the large deviations of the smallest eigenvalue.
The technique improves upon past methods by not requiring the explicit law of the eigenvalues, and we
apply it to a large class of random matrices that were previously out of reach.
In particular, we solve an open problem related to the performance of principal components analysis on highly correlated data,
and open the way towards analyzing the high-dimensional landscapes of complex inference models.
We probe our results using an importance sampling approach, effectively simulating events with probability as small as $10^{-100}$.
\end{abstract}

\maketitle

Theoretical physics and random matrix theory share a long history that dates back to Wigner \cite{wigner1955}, 
and that powered progress in various areas ranging from disordered systems \cite{edwards1975theory,sherrington1975solvable} to quantum chaos \cite{bohigas1984characterization},
quantum chromodynamics \cite{verbaarschot2000random}, or superconductivity \cite{bahcall1996random}.
The growing interplay of physics and statistics \cite{zdeborova2016statistical,gabrie2020mean,zdeborova2020understanding}
further strengthened this connection.
A textbook example of this bond is principal components analysis (PCA), a statistical estimation method based on random matrix theory, 
and applied in fields as diverse as image compression \cite{abdi2010principal,fukunaga2013introduction,du2007hyperspectral,zhang2010two}, neurosciences \cite{brenner2000adaptive,jirsa1994theoretical}, genetics \cite{reich2008principal}, or finance \cite{bouchaud2000theory}.

To fix our ideas, let $\bX \in \bbR^{m \times n}$ be the data matrix, whose columns $\{\bx_i\}_{i=1}^n$ are observations independently drawn from a Gaussian distribution $\mathcal{N}(0,\Gamma)$.
PCA aims at discovering a ``principal component'' eigenspace of the covariance matrix $\Gamma$ 
by studying the largest eigenvalue of the \emph{sample covariance matrix} $C_n \equiv \sum_i \bx_i \bx_i^\intercal/n$: indeed,
a strong outlier eigenvalue in $\Gamma$ typically induces a corresponding outlier in $C_n$ \cite{edwards1976eigenvalue,baik2005phase,benaych2011eigenvalues,aubin2020spiked}.

Pioneering physics works \cite{dean2006large,majumdar2009large} addressed the general question ``How good is PCA ?''. 
Precisely, they wished to understand if an outlier can appear in $C_n$ even if there is no structure to uncover in $\Gamma$:
this ``null hypothesis'' provides a model to gauge the significance of results obtained on a real-world dataset. 
Such atypical events are known as \emph{large deviations}, and the mentioned works, as well as subsequent ones, had to restrict to \emph{uncorrelated} data, in which $\Gamma$ is the identity matrix (or a finite-rank perturbation of it) \cite{dean2006large,majumdar2009large,biroli2020large,maida2007large,vivo2007large,majumdar2014top}.
Realistic data (\textit{e.g.}\ a natural image) indeed contain non-trivial correlations that the Coulomb gas analysis used in \cite{dean2006large,majumdar2009large}
is not equipped to handle.
While data structure is a key ingredient of learning and inference \cite{zdeborova2020understanding},
probing the statistical significance of PCA on correlated data remained an open question. 

The present letter addresses and solves this long-lasting problem for \emph{arbitrary} $\Gamma$, \textit{i.e.}\ PCA with correlated data. We further discuss other consequences of our results, 
notably in the physics of disordered systems.

\begin{figure}
\includegraphics[width=8.6cm]{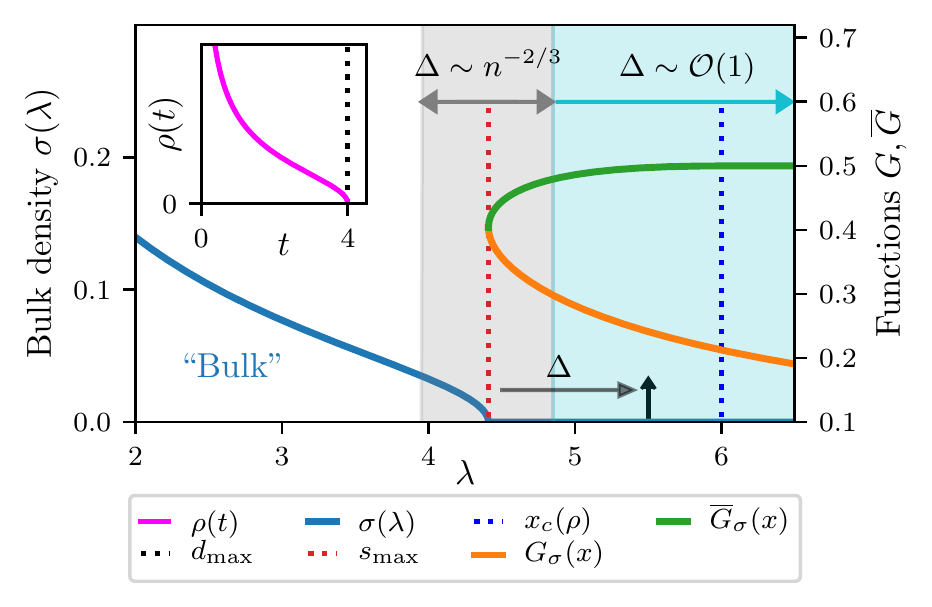}
\caption{The bulk $\sigma(\lambda)$, and the functions $G_\sigma,\overline{G}_\sigma$ for $\alpha = 2$ and $\rho(t)$ the Marchenko-Pastur law with ratio $1$. 
In the box, we plot $\rho(t)$ and the right edge $d_\mathrm{max}$ of its support. 
The black arrow is an outlier in the spectrum of $H_n$, and
$\Delta$ is the gap between this outlier and the bulk $\sigma(\lambda)$.
}
\label{fig:bulk_resolvent}
\end{figure}

To state our main result, we first define the mathematical quantities involved.
Letting $\bx_i = \sqrt{\Gamma} \bz_i$ with $\bz_i \sim \mathcal{N}(0,1)$, one can see that $C_n$ has the same eigenvalues (up to possible zeros and a scaling factor) as $H_n \equiv \bZ^\intercal \Gamma \bZ / m$.
The ``bulk'' of $H_n$, \textit{i.e.}\ the large $n$ limit of its eigenvalue density $(1/n) \sum_{i=1}^n \delta[\lambda - \lambda_i(H_n)]$, is denoted $\sigma(\lambda)$.
The large $n$ limit of the empirical spectral density of $\Gamma$ is denoted $\rho(t)$.
We show an example of $\sigma(\lambda)$ in Fig.~\ref{fig:bulk_resolvent}, when $\rho(t)$ is the Marchenko-Pastur law \cite{marchenko1967distribution}.
$\sigma(\lambda)$ is analytically derived using the \emph{Stieltjes transform} of random matrix theory:
$G_\sigma(x) = \mathrm{Tr}[(x-H_n)^{-1}]/n = \int \mathrm{d}\lambda \, \sigma(\lambda)/(x-\lambda)$. 
Assuming that $m/n \to \alpha$, the Marchenko-Pastur equation \cite{marchenko1967distribution} gives the inverse of $G_\sigma$ :
\begin{equation}\label{eq:MP_equation}
   G_\sigma^{-1}(\omega) = \frac{1}{\omega} + \alpha \int \mathrm{d}t \rho(t) \frac{t}{\alpha - t \omega}.
\end{equation}
$\sigma(\lambda)$ is determined by $G_\sigma(x)$ via the Stieltjes-Perron inversion formula $\sigma(\lambda) = -\lim_{\epsilon \downarrow 0}\mathrm{Im}[G_\sigma(\lambda+i \epsilon)] / \pi$.
In particular, the support of $\sigma(\lambda)$ and its right edge $s_\mathrm{max}$
can be computed (analytically or numerically) from eq.~\eqref{eq:MP_equation}.
By rotation invariance of $\bZ$, one can diagonalize $\Gamma$, i.e.\ assume $\Gamma = \mathrm{Diag}(\{d_\mu\})$, with all $d_\mu \geq 0$, which implies that:
\begin{equation}\label{eq:def_Hn}
    H_n \equiv \frac{1}{m} \sum_{\mu=1}^m d_\mu \bz_\mu \bz_\mu^\dagger, 
\end{equation}
in which the $\{\bz_\mu\}$ are standard Gaussian vectors. 
This leads us to further extend the matrix model to \emph{generalized sample covariance matrices}, in which the fixed variables $d_\mu$ of eq.~\eqref{eq:def_Hn} are not necessarily positive,
and the $\bz_\mu$ can be real or complex.
Importantly, the positivity (or negativity) of the matrix $H_n$ is equivalent to the positivity (or negativity) of all $d_\mu$.
We denote $\rho(t) = \lim_{m \to \infty} \sum_{\mu=1}^m \delta(t - d_\mu)/m$, and $d_\mathrm{max}$ the right edge of the support of $\rho(t)$, that we assume to be bounded (see the inner box in Fig.~\ref{fig:bulk_resolvent}).

In the following, we detail our main result before discussing its consequences, notably for an old open problem in the physics of disordered systems.
We then probe our findings using precise numerical simulations.
The remaining of the letter is devoted to the derivation of our result.

{\bf Large deviations -}
From now on we restrict to the study of $\lambda_\mathrm{max}(H_n)$. Since we can always consider $d'_\mu = -d_\mu$, our analysis also applies to $\lambda_\mathrm{min}(H_n)$.
We emphasize that the large deviations regime corresponds to \emph{macroscopic} changes in $\lambda_\mathrm{max}(H_n)$, which are exponentially rare, 
as opposed to the typical fluctuations, which are generically in the scale $n^{-2/3}$ \cite{tracy1994level,johnstone2001distribution,ding2020tracy}.
These two regimes are shown as cyan and grey regions in Fig.~\ref{fig:bulk_resolvent}.
Crucially, we assume that
$\max_{1 \leq\mu \leq m} d_\mu$ approaches $d_\mathrm{max}$ as $m \to \infty$, \textit{i.e.}\ that \emph{there is no outlier in the list $\{d_\mu\}$}.
This ensures that $\lambda_\mathrm{max}(H_n)$ converges to the right edge $s_\mathrm{max}$ of the bulk $\sigma(\lambda)$. 
In other words, the set of vectors $\{\bz_\mu\}$ such that the spectrum of $H_n$ has an outlier is
very atypical under the Gaussian distribution. 

We now state our result under the aforementioned hypotheses.
Let $\beta \in \{1,2\}$ for respectively real and complex $\bz_\mu$, with the convention $\langle |z|^2 \rangle = 1$ for a Gaussian standard random variable.
We denote $P_n(x)$ the PDF of $\lambda_\mathrm{max}(H_n)$ (for \emph{given} $\{d_\mu\}$).
For $x \geq s_\mathrm{max}$:
\begin{equation}\label{eq:ldp}
     P_n(x)  \simeq \exp \Big\{  -n \overbrace{\frac{\beta}{2} \int_{s_\mathrm{max}}^x    [\overline{G}_\sigma(u) - G_\sigma(u)] \mathrm{d}u}^{I(x)} \Big\}.
\end{equation}
The function $\overline{G}_\sigma$ is defined in the following (technical) way.
By monotonicity arguments, it can easily be seen that the equation $G_\sigma^{-1}(\omega) = x$ has a second solution $\omega = \overline{G}_\sigma(x)$, 
sometimes referred to as the ``second branch'' of the Marchenko-Pastur equation~\eqref{eq:MP_equation}.
Examples of $(G_\sigma,\overline{G}_\sigma)$ are given in Fig.~\ref{fig:bulk_resolvent}.
An important remark is that $\overline{G}_\sigma(x)$ can saturate if $d_\mathrm{max} > 0$ (\textit{i.e.}\ if $H_n$ is not negative).
In this case, $\overline{G}_\sigma(x) = \alpha / d_\mathrm{max}$ for $x \geq x_c(\rho)$, with
\begin{equation}\label{eq:def_xc}
   x_c(\rho) \equiv d_\mathrm{max}^2 G_\rho(d_\mathrm{max}) + (\alpha^{-1}-1) d_\mathrm{max}.
\end{equation}
Here, $G_\rho(z) = \int \mathrm{d}t \, \rho(t) / (z-t)$ is the Stieltjes transform of $\rho(t)$.
Possibly, $x_c(\rho) = +\infty$ if $G_\rho(d_\mathrm{max}) = +\infty$.
If $H_n$ is negative, then $\overline{G}_\sigma(x)$ diverges to $+\infty$ as $x \uparrow 0$, and we set $\overline{G}_\sigma(x) = +\infty$ for $x \geq 0$.

{\bf Discussion - }
Eq.~\eqref{eq:ldp} is the main result of this letter.
The negative of the argument of the exponential is called the \emph{rate function} $I(x)$ in the large deviations language.
\begin{figure}
\includegraphics[width=8.6cm]{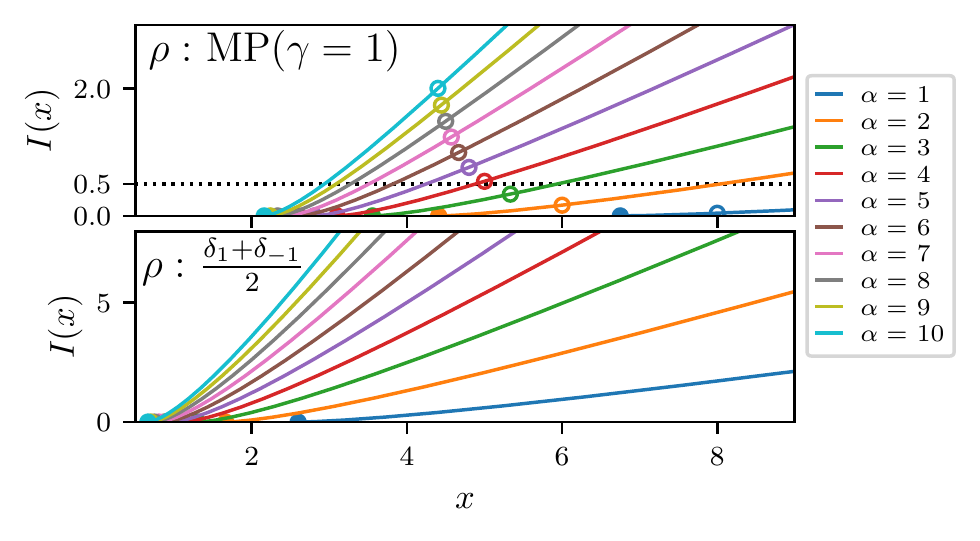}
\caption{The rate function $I(x)$ for different values of $\alpha$ and two different distributions $\rho$, in the real case. The full dots show the right edge $s_\mathrm{max}$ of the 
bulk, while the empty dots (when present) correspond to the transition $x_c(\rho)$.
}
\label{fig:rate_function}
\end{figure}
In Fig.~\ref{fig:rate_function}, we show analytical computations of $I(x)$ for different $\alpha$ and $\rho(t)$.

Our analytical large deviations computation also
paves the way toward a direct understanding of the topology of the landscape of complex inference models.
Indeed, it allows to investigate the number of local minima in these landscapes, using that $H_n$ is related to the Hessian matrix of complex statistical and disordered models, such as the perceptron \cite{maillard2020landscape}.
Exploring precisely these landscapes is an important open problem in the disordered systems community,
as the traditional methods have been limited to simpler Gaussian models \cite{ros2019complex}, 
and our results are an important step in this exciting direction.

{\it Consistency with previous results -}   
Importantly, in the Wishart case, i.e.\ $\rho(t) = \delta(t-1)$, our result is consistent with previous works \cite{majumdar2009large,biroli2020large}. 
Indeed, as detailed in Appendix~\ref{sec_app:verification_wishart}, eq.~\eqref{eq:ldp} reduces in this case to the known expression:
\begin{equation*}
  P_n(x) \simeq \exp \Big\{- \!\frac{\alpha n}{2}\! \int_{\lambda_+(\alpha)}^x\!\! \frac{\sqrt{(u - \lambda_+(\alpha))(u-\lambda_-(\alpha))}}{u} \! \mathrm{d}u \Big\},
\end{equation*}
with $\lambda_+(\alpha) \equiv (1 + \alpha^{-1/2})^2$.

{\it A phase transition -} 
Let us describe a first notable consequence of eq.~\eqref{eq:ldp}.
We assume that $d_\mathrm{max} > 0$ and that $x_c(\rho)$ (see eq.~\eqref{eq:def_xc}) is finite, 
\textit{e.g.}\ $\rho(t)$ can be the Marchenko-Pastur law, as shown in Figs.~\ref{fig:bulk_resolvent},\ref{fig:rate_function}.
Recall that $\overline{G}_\sigma(x)$ saturates at $\alpha/d_\mathrm{max}$ for $x \geq x_c(\rho)$. 
It is in general not smooth at $x = x_c(\rho)$ and
this singularity induces a phase transition in the rate function $I(x)$. The \emph{order} of the transition (\textit{i.e.}\ the order of the first discontinuous derivative of $I(x)$)
can be computed if the right tail of $\rho(t)$ behaves as $\rho(t) \sim (d_\mathrm{max} - t)^\eta$ for $t \to d_\mathrm{max}$,  with $\eta > 0$, so that $x_c(\rho) < \infty$.
When $\eta \geq 1$ and $1/2 \leq \eta < 1$ (\textit{e.g.}\ the Marchenko-Pastur law, for which $\eta = 1/2$) we show that the transition is respectively of second and third order.
The details are given in Appendix~\ref{sec_app:phase_transition}, and we conjecture generically the order to be $k + 1$ if $1/k  \leq \eta < 1/(k - 1)$.
We note that a similar argument based on the vanishing exponent of the density was already used in the literature, in the context of multi-critical matrix models \cite{majumdar2014top}.

{\bf Monte-Carlo simulations -}
Although eq.~\eqref{eq:ldp} is a large-$n$ result,
we investigate numerically the large deviations regime at moderately large $n$, which is the relevant regime for real data in PCA.
Because we need $(1/m) \sum_\mu \delta(t-d_\mu) $ to be very close to $\rho(t)$, we can not perform 
histograms of $\lambda_\mathrm{max}(H_n)$, as performed in \cite{majumdar2009large}, since the large deviations probability decays exponentially in $n$.
We instead modify the law of $\bz$ so that it favors large deviations, a technique which is known as \emph{importance sampling} \cite{bucklew2013introduction}.
This powerful Monte-Carlo method allows to numerically access the tails of a given high-dimensional probability distribution
and has been successfully applied to various problems across the physical sciences, 
from random graphs \cite{engel2004large} to simulations of the height distribution in the KPZ equation \cite{hartmann2018high},
and random matrices \cite{driscoll2007searching,saito2010multicanonical}, as in the present letter. 
For a more exhaustive description of the applications of importance sampling in physics, we refer the reader to \cite{hartmann2018high}.
Let us now detail the technicalities of the approach.

We denote $\mathcal{D}\bz \equiv \mathrm{d}\bz \ e^{-||\bz||^2/2} / (2\pi)^{n/2}$ the standard Gaussian law.
We will tilt this Gaussian distribution by explicitly giving more weight to configurations having a larger $\lambda_\mathrm{max}(H_n)$.
More precisely, we aim at sampling from the distribution
\begin{equation*}
   P_t(\bz)\mathrm{d}\bz \propto \mathcal{D} \bz \, e^{n t \lambda_\mathrm{max}(H_n)}.
\end{equation*}
For a given $t \geq 0$, eq.~\eqref{eq:ldp} and the Laplace method imply that when sampling $\bz$ under $P_t$, the largest eigenvalue of $H_n$ concentrates on 
$x^\star(t) \equiv \argmin_x [t x - I(x)]$.
One sees clearly now that sampling from this tilted distribution gives information about the Legendre transform of the large deviations function $I(x)$.

We implement a Metropolis-Hastings algorithm in order to sample from $P_t$.
The physical parameters are $n,m,\rho,t$, and we generate i.i.d.\ samples $\{d_\mu\}_{\mu=1}^m$ from $\rho$. 
We initialize $\{\bz_\mu\}_{\mu=1}^m$ as standard Gaussian vectors, and sample from the move proposal distribution 
$g(\bz' | \bz)$ as follows:
\\
$(i)$ Pick a random index $\mu$ with probability $P(\mu) \propto e^{\beta_d d_\mu}$.
\\
$(ii)$ Draw a uniform $\bbe \in \bbR^n$ with norm $||\bbe||^2 = n$, 
and draw $L \geq 0$ from a truncated Gaussian distribution centered in $1$ and with variance $\Delta > 0$. 
Let $\bz'_\mu = \sqrt{L} \bbe$.
\\
$(iii)$ The new state $\bz'$ is given by changing $\bz_\mu \to \bz'_\mu$.

\begin{figure}
\includegraphics[width=8.6cm]{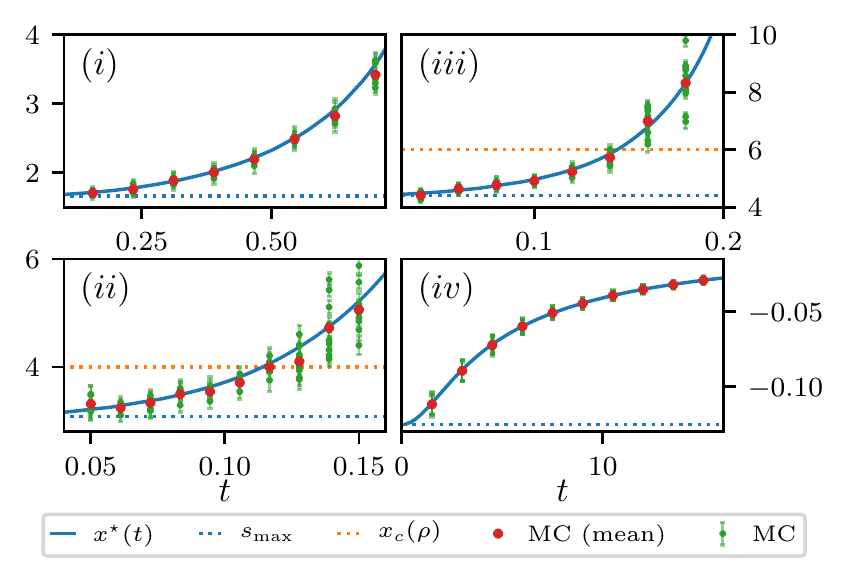}
\caption{The function $x^\star(t)$ for $\rho(t) = $ : $(i)$ the sum $(\delta_{1}+\delta_{-1})/2$, $(ii)$ Wigner's semicircle law, $(iii)$ the Marchenko-Pastur law with ratio $1$, $(iv)$ the uniform distribution in $[-2,-1]$.
In all cases $\alpha = 2$ except for $(ii)$, in which $\alpha = 1$.
Solid lines are analytical predictions. The different Monte-Carlo runs ($n = 500$) are shown in green with their respective noise. The mean of the green points 
is depicted as a red dot.}
\label{fig:mc}
\end{figure}
We impose the detailed balance condition with stationary distribution $P_t(\bz)$
and move proposal distribution $g(\bz'|\bz)$ in the MCMC.
We measure the largest eigenvalue of $H_n$, which we compare to $x^\star(t)$.
The parameters $(\beta_d,\Delta)$ are found to reduce greatly the equilibration time of the Markov chain, 
and are adapted during a warmup phase to obtain an acceptance
ratio in the range $[0.2,0.3]$. Physically, $\beta_d$ favors the right edge of the bulk, while $\Delta$ favors 
large norms of $\bz_\mu$. 
The code is available in a public \href{https://github.com/AnMaillard/LD_lmax_sample_covariance}{repository} \cite{github_repo} and
the results of the simulations are given in Fig.~\ref{fig:mc} for four choices of $\rho(t)$.
The agreement with our predictions is excellent, whether $H_n$ is negative, positive, or neither.
Even though the variability of the results naturally increases with $t$, 
we are able to access very large values of $x^\star(t)$, beyond the transition point $x_c(\rho)$.
For example, in case $(iii)$ of Fig.~\ref{fig:mc} we sample up to $x^\star(t) \simeq 8$.
Comparing with Fig.~\ref{fig:rate_function}, this implies that our importance sampling simulations reach events with probability of order $e^{-0.5 n} \sim 10^{-109}$ under a naive sampling.

{\bf Derivation of the result -}
Let us now derive eq.~\eqref{eq:ldp}, focusing on the real case.
Our derivation is based on a \emph{tilting} method, developed in a series of recent mathematical works \cite{biroli2020large,augeri2019large,mckenna2019large,guionnet2020large,husson2020large,belinschi2020large}.
This technique is more adaptable than a Coulomb gas analysis, as it does not 
require the joint probability of the eigenvalues of $H_n$, which is not known here.
Moreover, the calculation does not rely on any heuristics, and we expect it to be adaptable
into mathematically rigorous statements.

{\it Informal introduction to the method -}
To fix the ideas, we let $x \geq s_\mathrm{max}$, and we aim at computing $P_n(x)$, i.e.\ the probability of a rare event 
in which $\lambda_\mathrm{max}(H_n)$ is close to $x$ rather than its typical value $s_\mathrm{max}$.
The main idea of the method is to \emph{tilt} the probability density of $H_n$ so that having $\lambda_\mathrm{max}(H_n) \simeq x$
becomes a \emph{typical} event, rather than a rare one.
More precisely, this new tilted law will be parametrized by a number $\theta \geq 0$:
for each $\theta$, the largest eigenvalue will typically be close to a value $x(\theta)$ for large $n$.
Conversely, each $x \geq s_\mathrm{max}$ will be associated to a $\theta_x \geq 0$, and a tilting parametrized by $\theta_x$ will typically induce the largest eigenvalue to be 
close to $x$. As we will see, we will gain access to $P_n(x)$ by studying this function $\theta_x$.

{\it Tilting the measure -}
We start with a simple use of the tilting method\footnote{The analysis of \cite{biroli2020large,belinschi2020large} suggests a tilting which is function of $A_n$, with $H_n=A_n A_n^\intercal$. 
However for arbitrary $d_\mu$, $A_n$ is not defined so that we use this simpler tilting.
We shall later come back to this idea by allowing complex-valued $A_n$.}.
The simplest possible tilting of the measure, inspired by the aforementioned mathematical works, is an exponential tilting.
More precisely, we define the tilted distribution of $\bz$ as:
\begin{equation}\label{eq:distribution_tilted_first}
   P_{\theta,\bbe}(\bz) \mathrm{d}\bz \propto \mathcal{D}\bz \ e^{\frac{\theta n}{2} \bbe^\intercal H_n \bbe},
\end{equation}
for a given vector $\bbe$ (such that $||\bbe||^2 = 1$) and a parameter $\theta \geq 0$.
Recall that $\mathcal{D}\bz \equiv \mathrm{d}\bz \ e^{-||\bz||^2/2} / (2\pi)^{n/2}$ is the standard Gaussian law.
As we will see, this tilting induces a macroscopic move of the largest eigenvalue that only depends on $\theta$.
Note that the tilted distribution $P_{\theta,\bbe}$ is equivalent to a rank-one change in the covariance of the $\{\bz_\mu\}$.
Indeed, using simple algebra detailed in Appendix~\ref{subsec_app:tilt_Hn} we show that, under the tilted law of eq.~\eqref{eq:distribution_tilted_first}, $H_n$ is distributed as:
\begin{equation*}
     H_n^{(\bbe,\theta)} = \frac{1}{m} \sum_{\mu=1}^m d_\mu [\mathds{1}_n  +  \kappa_\theta(d_\mu) \bbe \bbe^\intercal] \bz_\mu \bz_\mu^\intercal [\mathds{1}_n  +  \kappa_\theta(d_\mu) \bbe \bbe^\intercal],
\end{equation*}
with $\kappa_\theta(t) \equiv (1- \alpha^{-1}\theta t)^{-1/2}  - 1$. When $n \to \infty$, the largest eigenvalue of $H_n^{(\bbe,\theta)}$ typically approaches 
a value $x(\theta) \geq s_\mathrm{max}$:
since $H_n^{(\bbe,\theta)}$ is a finite-rank change of $H_n$, its largest eigenvalue can indeed be typically larger than $s_\mathrm{max}$. 
Moreover, we see from the expression of $H_n^{(\bbe,\theta)}$ that as $\theta \to \alpha/d_\mathrm{max}$, some of the coefficients $\kappa_\theta(d_\mu)$ 
will grow very large: we thus expect that for sufficiently large $\theta$, such an outlier eigenvalue will indeed pop out of the ``bulk''.
Note that $x(\theta)$ does not depend on the direction of $\bbe$, as the Gaussian distribution of the vectors $\bz_\mu$ is 
rotationally invariant.

Let us now see how to relate $P_n(x(\theta))$ to this tilted distribution. We can write the trivial identity
\begin{eqnarray}\label{eq:tilting_first}
    &&P_n(x(\theta)) = \int \mathcal{D} \bz \ \delta(\lambda_\mathrm{max}(H_n) - x(\theta)), \nonumber \\
    &&= \int \mathcal{D}\bz \ \delta(\lambda_\mathrm{max}(H_n) - x(\theta)) \ \frac{\int_{\norm{\bbe}^2 = 1} \mathrm{d}\bbe \ e^{\frac{\theta n}{2} \bbe^\intercal H_n \bbe}}{\int_{\norm{\bbe}^2 = 1} \mathrm{d}\bbe \ e^{\frac{\theta n}{2} \bbe^\intercal H_n \bbe}}, \nonumber \\
    &&=\int_{\norm{\bbe}^2 = 1} \hspace{-0.5cm} \mathrm{d}\bbe \ \mathcal{D}\bz \ \delta(\lambda_\mathrm{max}(H_n) - x(\theta)) \frac{e^{\frac{\theta n}{2} \bbe^\intercal H_n \bbe}}{e^{n J_n(H_n,\theta)}}.
\end{eqnarray}
We introduced $J_n(H_n,\theta) \equiv (1/n) \ln \int_{\norm{\bbe}^2 = 1}\mathrm{d}\bbe \ e^{\frac{n \theta}{2} \bbe^\intercal H_n \bbe}$.
In eq.~\eqref{eq:tilting_first}, we almost see the tilted probability distribution of eq.~\eqref{eq:distribution_tilted_first} appearing.
However this is not exactly the case, as the term $e^{n J_n(H_n,\theta)}$ also depends on $\bz$.
If this term would not depend on $\bz$ at leading exponential order in $n$, the law of $\bz$ would be the tilted law of eq.~\eqref{eq:distribution_tilted_first}, and we could remove the $\delta$ term in eq.~\eqref{eq:tilting_first}:
indeed, the constraint $\lambda_\mathrm{max}(H_n) \simeq x(\theta)$ would already be satisfied, by the very definition of $x(\theta)$ !

{\it The spherical integrals -}
Therefore, we study first $J_n(H_n,\theta)$.
Let us introduce $J_1(\theta,x)$, defined as the limit of
$J_n(H_n,\theta)$, \emph{assuming} $\lambda_\mathrm{max}(H_n) \to x$ as $n \to \infty$ (which we can safely assume because of the constraint in eq.~\eqref{eq:tilting_first}).
Adopting the language of statistical physics, we call $J_1$ a \emph{quenched} spherical integral. 
More precisely, $J_1$ belongs to a class of high-dimensional integrals known as Harish-Chandra-Itzykson-Zuber (HCIZ) integrals \cite{harish1957differential,itzykson1980planar}.
To compute $J_1(\theta,x)$, we introduce a Lagrange multiplier to fix the norm of $\bbe$. 
This yields:
\begin{align*}
   J_n(H_n,\theta) &= \frac{1}{n} \ln \frac{\int \mathrm{d}\bbe \, \delta(||\bbe||^2 - 1) \, e^{\frac{\theta n}{2} \bbe^\intercal H_n \bbe}}{\int \mathrm{d}\bbe \, \delta(||\bbe||^2 - 1)}\\ 
   & =\extr_{\gamma}\Big[\frac{\gamma}{2} + \frac{1}{n} \ln \int \mathrm{d}\bbe \ e^{- \frac{\gamma}{2} ||\bbe||^2 + \frac{\theta n}{2} \bbe^\intercal H_n \bbe}\Big] \\ 
   & \hspace{1cm}-  \extr_{\gamma}\Big[\frac{\gamma}{2} + \frac{1}{n} \ln \int \mathrm{d}\bbe \ e^{- \frac{\gamma}{2} ||\bbe||^2}\Big], \\ 
   & =\extr_{\gamma}\Big[\frac{\gamma}{2}  - \frac{1}{2n} \mathrm{Tr} \log (\gamma \mathds{1}_n - \theta H_n)\Big] - \frac{1}{2}.
\end{align*}
Here, $\extr_\gamma f(\gamma)$ means solving the saddle-point equation $\partial_\gamma f(\gamma) = 0$.
Importantly, the Lagrange multiplier $\gamma$ must be such that the matrix $\gamma - \theta H_n$ is definite positive, for the Gaussian integral to be well-defined: since we assumed that $\lambda_\mathrm{max}(H_n) \! \to \! x$, 
this implies that $\gamma > \theta x$.
Moreover, it is easy to see by monotonicity arguments that this extremum is actually an infimum. 
All in all, we arrive at:
\begin{equation*}
   J_n(H_n,\theta) = \inf_{\gamma > \theta x} \Big[\frac{\gamma}{2} - \frac{1}{2n} \mathrm{Tr} \log (\gamma \mathds{1}_n - \theta H_n)\Big] - \frac{1}{2}.
\end{equation*}
From this expression, we can see easily that if $\sigma(u)$ is the asymptotic spectral density of $H_n$, we have:
\begin{equation}\label{eq:J1_variational}
   J_1(\theta,x) = \inf_{\gamma > \theta x} \big[\frac{\gamma}{2} - \frac{1}{2} \int \mathrm{d}u \, \sigma(u) \ln (\gamma - \theta u)\big] - \frac{1}{2}.
\end{equation}
Let us come back to eq.~\eqref{eq:tilting_first}. Using what we just described, we have, at leading exponential order:
\begin{align}\label{eq:Pn_1}
   P_n(x(\theta)) &\simeq \int_{\norm{\bbe}^2 = 1} \hspace{-0.5cm} \mathrm{d}\bbe \ \mathcal{D}\bz \ \delta(\lambda_\mathrm{max}(H_n) - x(\theta)) \frac{e^{\frac{\theta n}{2} \bbe^\intercal H_n \bbe}}{e^{n J_1(\theta,x(\theta))}}, \nonumber \\
   &\simeq e^{-n J_1(\theta,x(\theta))} \int_{\norm{\bbe}^2 = 1} \hspace{-0.5cm} \mathrm{d}\bbe \ \mathcal{D}\bz \ e^{\frac{\theta n}{2} \bbe^\intercal H_n \bbe}.
\end{align}
As already argued, we removed the $\delta$ constraint in eq.~\eqref{eq:Pn_1} by definition of $x(\theta)$: under the tilted law $P_{\theta,\bbe}$, 
the largest eigenvalue $\lambda_\mathrm{max}(H_n)$ typically concentrates on $x(\theta)$, so this constraint is naturally satisfied.
The expression of eq.~\eqref{eq:Pn_1} involves another integral, that we call \emph{annealed} and denote $F_n(\theta)$, borrowing again from the statistical physics jargon:
\begin{equation}\label{eq:def_Fn}
   F_n(\theta) \equiv \frac{1}{n}\ln \int \mathcal{D}\bz \int_{\norm{\bbe}^2 = 1} \mathrm{d}\bbe \ e^{\frac{\theta n}{2} \bbe^\intercal H_n \bbe}.
\end{equation}
Similarly to $J_n$, we denote by $F_1(\theta)$ the limit of $F_n(\theta)$.
If $d_\mathrm{max} > 0$, we also impose $\theta < \alpha/d_\mathrm{max}$ so that $F_n(\theta)$ is well-defined.
We compute it by direct integration on $\bz$ in eq.~\eqref{eq:def_Fn}, since $H_n = (1/m) \sum_\mu d_\mu \bz_\mu \bz_\mu^\intercal$:
\begin{equation}\label{eq:F1_explicit}
   F_1(\theta) = -\frac{\alpha}{2} \int \mathrm{d}t \rho(t) \ln (1 - \alpha^{-1} \theta t) .
\end{equation}
Combined with eq.~\eqref{eq:Pn_1}, this implies
\begin{equation}\label{eq:Pn_2}
   P_n(x(\theta)) \simeq \exp\{-n [J_1(\theta,x(\theta)) - F_1(\theta)]\}.
\end{equation}
Note that we imposed $\theta < \theta_\mathrm{max}$, with $\theta_\mathrm{max} \equiv \alpha / d_\mathrm{max}$ if $d_\mathrm{max} > 0$ and $+ \infty$ otherwise.
Conversely, this implies that eq.~\eqref{eq:Pn_2} can only be applied for $s_\mathrm{max} \leq x < x_\mathrm{max}$, 
with $x_\mathrm{max} = x(\theta_\mathrm{max})$ (which can be $+ \infty$).
This creates a possibly important limitation of the tilting we used, if $x_\mathrm{max}$ is finite: in this case, the method does not give access to the large deviations for $x \geq x_\mathrm{max}$!
We will precisely characterize when such a limitation occurs in the following, relating it to the phase transition phenomenon described above, and we will develop a different tilting to circumvent this issue.

{\it Simplifying the rate function -}
First, let us focus on $x < x_\mathrm{max}$ and show that we find eq.~\eqref{eq:ldp}. 
We can rewrite eq.~\eqref{eq:Pn_2} as:
\begin{equation}\label{eq:Pn_3}
P_n(x) \simeq \exp\{-n [J_1(\theta_x,x) - F_1(\theta_x)]\}.
\end{equation}
Recall that $\theta_x$ is chosen exactly to be able to remove the delta constraint in eq.~\eqref{eq:Pn_1}. 
However, for any $\theta' \geq 0$, we can always write an equivalent to eq.~\eqref{eq:Pn_1}, keeping the delta constraint:
\begin{equation*}
   P_n(x) \simeq \int_{\norm{\bbe}^2 = 1} \hspace{-0.5cm} \mathrm{d}\bbe \ \mathcal{D}\bz \ \delta(\lambda_\mathrm{max}(H_n) - x) \frac{e^{\frac{\theta' n}{2} \bbe^\intercal H_n \bbe}}{e^{n J_1(\theta',x)}}.
\end{equation*}
From here, we can always upper-bound $P_n(x)$ by simply discarding the delta constraint in this equation, which gives
$P_n(x) \lessapprox e^{-n[J_1(\theta',x)-F_1(\theta')]}$ at leading exponential order. Combining this with eq.~\eqref{eq:Pn_3}, we see that we can write (recall $P_n(x) \simeq e^{-n I(x)}$):
\begin{equation}\label{eq:Pn_final}
 I(x) = \sup_\theta[J_1(\theta,x) - F_1(\theta)].
\end{equation}
We focus now on simplifying the rate function of eq.~\eqref{eq:Pn_final}, to obtain eq.~\eqref{eq:ldp}. 
We need to study the behavior of the quenched integral $J_1$ of eq.~\eqref{eq:J1_variational}. This type of integral has been studied 
in the context of $2$-spin glass models, for different spectra $\sigma(\lambda)$, in the physics and mathematics literature \cite{kosterlitz1976spherical,marinari1994replica,parisi1995mean,guionnet2005fourier,benaych2011rectangular}.
We recall here known results on $J_1(\theta,x)$, stated for instance in \cite{maillard2019high}.
Cancelling the derivative with respect to $\gamma$ in eq.~\eqref{eq:J1_variational} yields the equation:
\begin{equation}
   \int \mathrm{d}u \, \frac{\sigma(u)}{\gamma - \theta u} = 1.
\end{equation}
This equation is solved by $\gamma^\star = \theta G_\sigma^{-1}(\theta)$.
Plugging back this solution in eq.~\eqref{eq:J1_variational}, and using eq.~\eqref{eq:MP_equation} yields that $J_1(\theta,x) = F_1(\theta)$. 
We give more details on this in Appendix~\ref{subsec_app:simplifiying_J1}.
However, note that $\gamma$ is constrained to be smaller than $\theta x$. 
Therefore, the infimum in eq.~\eqref{eq:J1_variational} is reached in $\gamma^\star$ only for $\theta \leq \theta_c(x)$, with $\theta_c(x) \equiv G_\sigma(x)$.
At $\theta = \theta_c(x)$, $J_1(\theta,x)$ undergoes a transition, as $\gamma$ ``saturates'' at its limit value $\theta x$ for $\theta \geq \theta_c(x)$.
All in all, we reach:
\begin{eqnarray}\label{eq:J1_expression}
   &&J_1(\theta,x)= \\
   && \nonumber
   \begin{cases}
        F_1(\theta) = -\frac{\alpha}{2} \int \mathrm{d}t \rho(t) \ln (1 - \alpha^{-1} \theta t) &\mathrm{if} \ \theta \leq G_\sigma(x), \\    
        \frac{\theta x - 1 - \ln \theta}{2} - \frac{1}{2}\int \mathrm{d}u \ \sigma(u) \ln(x - u) &\mathrm{if} \ \theta \geq G_\sigma(x).
   \end{cases} 
\end{eqnarray}
Using eq.~\eqref{eq:J1_expression} in the result of eq.~\eqref{eq:Pn_final}, it is simple algebra to see that 
the maximum of $J_1(\theta,x) - F_1(\theta)$ is reached in $\theta_x = \overline{G}_\sigma(x)$.
Differentiating the resulting expression yields $I'(x) = (1/2) [\overline{G}_\sigma(x)-G_\sigma(x)]$, which gives 
eq.~\eqref{eq:ldp} and solves the problem in this case. We defer these algebraic details to Appendix~\ref{subsec_app:simplify_rate_dmax_negative}.

{\it Limitations of the tilting -}
As we mentioned, our method is not capable of predicting the large deviations for $x \geq x_\mathrm{max} = x(\theta_\mathrm{max})$.
Since we showed that $\theta_x = \overline{G}_\sigma(x)$, we can separate two cases:
\begin{itemize}[leftmargin=*]
   \item If $d_\mathrm{max} \leq 0$, then $\theta_\mathrm{max} = +\infty$ by definition, and therefore $x_\mathrm{max} = 0$, as we showed $\lim_{x \uparrow 0} \overline{G}_\sigma(x) = +\infty$ 
   below eq.~\eqref{eq:ldp}. For $x \geq 0$, $\overline{G}_\sigma(x) = +\infty$ and so eq.~\eqref{eq:ldp} is valid (indeed $I(x) = +\infty$ since $H_n$ is negative).
   In the end, our tilting allowed to compute the large deviations rate function $I(x)$ for any $x \geq s_\mathrm{max}$ in this case.
   \item  If $d_\mathrm{max} > 0$, then $\theta_\mathrm{max} = \alpha / d_\mathrm{max}$. 
   Since $\overline{G}_\sigma(x_c(\rho)) = \alpha / d_\mathrm{max}$, this
   yields that $x_\mathrm{max} = x_c(\rho)$, given by eq.~\eqref{eq:def_xc}.
   Therefore, we see that in this case, the condition for the tilting to be able to induce arbitrarily large outliers is $x_c(\rho) = +\infty$, i.e.\ $G_\rho(d_\mathrm{max}) = +\infty$.
   As we saw, the finiteness of $G_\rho(d_\mathrm{max})$ is exactly the existence condition of a phase transition in $I(x)$, which prevents the tilting from capturing all the large deviations.
\end{itemize}

{\it Beyond the transition -}
Here, we briefly outline the method we use to go beyond the phase transition when $d_\mathrm{max} > 0$, to circumvent the limitation described above.
As the method is extremely similar to the one we just described in detail, we focus only on the main steps and quantities.
We change the tilt of eq.~\eqref{eq:distribution_tilted_first} to (recall that $\mathcal{D}\bz$ is the standard Gaussian law):
\begin{equation}\label{eq:tilted_distribution_2}
    P_{\theta,\bbe,\bbf}(\bz) \, \mathrm{d}\bz \propto \mathcal{D}\bz \ e^{\frac{\theta n}{\sqrt{m}} \sum\limits_{i,\mu} \sqrt{d_\mu} e_i z_{\mu i} f_\mu},
\end{equation}
with $\sum_i e_i^2 = \sum_\mu  f_\mu^2 = 1$.
When $d_\mu \leq 0$, we define $\sqrt{d_\mu} \equiv i \sqrt{-d_\mu}$ so that the tilt is possibly complex-valued.
Eq.~\eqref{eq:tilted_distribution_2} corresponds to a simple additive shift of $\bz_\mu$, and the tilted law of $H_n$ is:
\begin{eqnarray*}
    H_n^{(\theta,\bbe,\bbf)} &\equiv& \frac{1}{m} \sum_{\mu=1}^m \Big[d_\mu \bz_\mu \bz_\mu^\intercal  + \frac{\theta^2 m}{\alpha^2} d_\mu^2 f_\mu^2 \bbe \bbe^\intercal
    \\
     &+& \frac{\theta \sqrt{m}}{\alpha} \mathds{1}_{\{d_\mu \geq 0\}} d_\mu^{3/2} f_\mu (\bbe \bz_\mu^\intercal  + \bz_\mu \bbe^\intercal) \Big]. \nonumber
\end{eqnarray*}
Let us give an intuitive view of the reasons why this tilting manages to induce the largest eigenvalue to be typically close to $x$, for 
any $x \geq s_\mathrm{max}$.
When $\theta = 0$ the largest eigenvalue of the unspiked matrix will naturally concentrate on $s_\mathrm{max}$.
As $\theta \gg 1$, a spike proportional to $\theta^2$ will push the largest eigenvalue of $H_n^{(\theta,\bbe,\bbf)}$ to $+\infty$.
By continuously varying $\theta$, this implies that the tilt can induce any outlier $x \geq s_\mathrm{max}$ in the spectrum.

The annealed and quenched ``HCIZ'' integrals corresponding to this tilting are:
\begin{eqnarray*}
   F_2(\theta) &=& \frac{1}{n} \ln \int \mathcal{D}\bz \int_{\norm{\bbe}^2 = 1} \hspace{-0.4cm} \mathrm{d}\bbe \int_{\norm{\bbf}^2 = 1} \hspace{-0.4cm}\mathrm{d}\bbf  \ e^{\frac{\theta n}{\sqrt{m}} \sum\limits_{i,\mu} \sqrt{d_\mu} e_i z_{\mu i} f_\mu}, \\*
   J_2(\theta,x) &=& \frac{1}{n} \ln \int_{\norm{\bbe}^2 = 1} \hspace{-0.4cm} \mathrm{d}\bbe \int_{\norm{\bbf}^2 = 1} \hspace{-0.4cm}\mathrm{d}\bbf  \ e^{\frac{\theta n}{\sqrt{m}} \sum\limits_{i,\mu} \sqrt{d_\mu} e_i z_{\mu i} f_\mu}.
\end{eqnarray*}
In $J_2(\theta,x)$, we assume that $\lambda_\mathrm{max}(H_n)$ converges to $x$ as $n \to \infty$. 
Introducing Lagrange multipliers in the spherical integrals, we find:
\begin{equation*}
   F_2(\theta) = \frac{\alpha}{2} \inf_{\gamma \geq d_\mathrm{max}} \hspace{-0.1cm} \Big[\frac{\gamma \theta^2}{\alpha^2} - \int \mathrm{d}t\rho(t) \ln(\gamma-t) - 1 - \ln\frac{\theta^2}{\alpha^2}\Big].
\end{equation*}
Similarly to our previous analysis of $J_1$, we show that there is a transition in $J_2$: for $\theta \leq \theta_c(x)$, $J_2(\theta,x) = F_2(\theta)$, while for $\theta \geq \theta_c(x)$ one reaches:
\begin{eqnarray*}
   &J_2(\theta,x) = \frac{\alpha-1}{2} \ln \big[\frac{1-\alpha + \sqrt{(\alpha-1)^2+4x\theta^2}}{2x}\big] - \frac{1+\alpha}{2}\\
   &- \frac{\alpha}{2} \ln \frac{\theta^2}{\alpha} + \frac{1}{2}\sqrt{(\alpha-1)^2+4x\theta^2} - \frac{1}{2}\!\int\!\mathrm{d}\lambda \,\sigma(\lambda) \ln (x - \lambda),
\end{eqnarray*}
with $\theta_c(x) \equiv \sqrt{x G_\sigma(x)^2 + (\alpha-1)G_\sigma(x)}$. 
The details of the derivations of $F_2$ and $J_2$ are given in Appendix~\ref{subsec_app:derivation_transition_F2_J2}.
Importantly, the very existence of the transition in $J_2(\theta,x)$ relies on the positivity of $x$, so that this tilting fails for negative matrices.
This notably implies that the tilt of eq.~\eqref{eq:distribution_tilted_first} is still crucial when $d_\mathrm{max} \leq 0$.

We deduce from the tilting method that $P_n(x)\!\simeq\!\exp\{ - n \sup_{\theta} [J_2(\theta,x) - F_2(\theta)]\}$ in the same way as before.
Using eq.~\eqref{eq:MP_equation} and the explicit expressions of $F_2$ and $J_2$ we derived, one shows that for all $x \geq s_\mathrm{max}$ the supremum is attained in $\theta_x = [x \overline{G}_\sigma(x)^2 + (\alpha-1) \overline{G}_\sigma(x)]^{1/2}$.
We compute then $I'(x) = [\overline{G}_\sigma(x) - G_\sigma(x)]/2$, which, together with $I(s_\mathrm{max}) = 0$, implies eq.~\eqref{eq:ldp}.
These algebraic calculations are detailed in Appendix~\ref{subsec_app:simplify_rate_dmax_positive}.
This ends the derivation of eq.~\eqref{eq:ldp} in all cases.

{\it A remark on the complex case -} 
We give an intuitive remark on how the factor $\beta = 2$ in the complex case of eq.~\eqref{eq:ldp} arises.
As we showed above, the method allows to write the large deviations function in the form $P_n(x) \simeq \exp\{-n\sup_{\theta}[J(\theta,x) - F(\theta)]\}$, with $F$ and $J$ annealed and quenched spherical integrals.
This result straightforwardly transfers to the complex setting, however the integrals $F$ and $J$ are now defined over unit vectors on the \emph{complex} unit sphere, i.e.\
they satisfy $\sum_i |e_i|^2 = 1$. 
It is known that the asymptotic behavior of these real and complex spherical integrals only differ by a factor $2$ (i.e.\ the complex integral is twice the real one), 
a phenomenon known as ``Zuber's $1/2$-rule''\cite{zinn2003some}: this explains the origin of the $\beta$ factor in eq.~\eqref{eq:ldp}.

{\bf The left tail of the large deviations -}
Importantly, we do not consider large deviations at the left of $s_\mathrm{max}$.
Such an event requires moving the whole bulk of eigenvalues, i.e.\ a number $\mathcal{O}(n)$ of eigenvalues, an event which has probability in the scale $\exp\{-n^2\}$ \cite{dean2006large,vivo2007large,majumdar2009large}.
Whether the method applied here could be extended to study this left tail is an interesting open question. 
As we saw, the core of the method is to create a tilt of the measure such that the largest eigenvalue is shifted in a controllable manner:
in this case, the tilting would need to induce a shift of the whole spectrum.
The perhaps most natural extension of the tilting of eq.~\eqref{eq:distribution_tilted_first} 
to this setting would be to consider an extensive-rank change in the covariance of the $\bz_\mu$:
\begin{equation*}
   \mathcal{D}\bz \to \mathcal{D}\bz \ e^{\frac{n}{2} \mathrm{Tr}[M_n \bO H_n \bO^\intercal]},
\end{equation*}
with $\bO$ an orthogonal matrix and $M_n$ an arbitrary matrix (with extensive rank) that will parametrize the tilting, similarly to $\theta$. 
Provided the mechanisms of the method we presented transfer to this case, this would give the large deviations function in terms of involved ``HCIZ'' spherical integrals.
The study of these extensive-rank HCIZ integrals in the high-dimensional limit was conducted in \cite{matytsin1994large}, and rigorously proven in 
\cite{guionnet2002large}.
The resulting formulas are however very involved, and the analysis of the left tail is thus left for future work.

{\bf Conclusion - }
We presented a generic technique to derive the right tail of the large deviations of the largest eigenvalues of random matrices. 
By symmetry, this also transfers to the left tail of the large deviations of the smallest eigenvalue.
This significantly improves over the seminal works of \cite{dean2006large,majumdar2009large}, solves a long-lasting open problem in statistics, and has deep consequences 
in the physics of disordered systems.
Thanks to the relative simplicity of our main result, we will further investigate its consequences in the future, in particular for PCA on real-world datasets, and for the 
landscape complexity of disordered systems.

{\bf Acknowledgments - }
The author is grateful to G.Biroli, F.Krzakala and S.Goldt for discussions, help and comments, and to A.Guionnet for introducing him to the tilting method used here.
Funding is acknowledged from ``Fondation CFM pour la Recherche''.


%

\newpage
\setcounter{secnumdepth}{2} 
\onecolumngrid
\appendix
\begin{center}
 \Large SUPPLEMENTARY MATERIAL
\end{center}

\section{Verification in the Wishart case}\label{sec_app:verification_wishart}

In the white Wishart case, we have $\rho(t) = \delta(t-1)$, and the density $\sigma(\lambda)$ is explicitly known, it is the Marchenko-Pastur distribution \cite{marchenko1967distribution}:
\begin{align}\label{eq:density_MP}
    \sigma(\lambda) &= \frac{\alpha}{2 \pi} \frac{\sqrt{(\lambda_+(\alpha)-\lambda) (\lambda - \lambda_-(\alpha))}}{\lambda} \mathds{1}\{\lambda_-(\alpha) \leq \lambda \leq \lambda_+(\alpha)\},
\end{align}
with $\lambda_\pm(\alpha) \equiv (1 \pm \alpha^{-1/2})^2$.
One can also explicitly solve eq.~(1) of the main text (which is just a quadratic equation in this case), 
and obtains for $x \geq \lambda_+(\alpha)$:
\begin{subnumcases}{}
 G_\sigma(x) = \frac{1-\alpha + \alpha x - \alpha \sqrt{(x - \lambda_+(\alpha))(x-\lambda_-(\alpha))}}{2 x}, & \\
 \overline{G}_\sigma(x) = \frac{1-\alpha + \alpha x + \alpha \sqrt{(x - \lambda_+(\alpha))(x-\lambda_-(\alpha))}}{2 x}. &
\end{subnumcases}
This implies that the rate function $I(x)$ (such that $P_n(x) \simeq e^{-n I(x)}$) satisfies, for every $x \geq \lambda_+(\alpha)$:
\begin{align}\label{eq:rate_MP}
    I(x) &= \frac{\alpha}{2} \int_{\lambda_+(\alpha)}^x \frac{\sqrt{(u - \lambda_+(\alpha))(u-\lambda_-(\alpha))}}{u} \ \mathrm{d}u.
\end{align}
On the other hand, the direct classical calculation using the joint law of eigenvalues of a Wishart matrix gives the following expression of the rate function 
(reminded for instance in Theorem~2.3 of \cite{biroli2020large}), for $x \geq \lambda_+(\alpha)$:
\begin{align}\label{eq:rate_MP_2}
    I(x) = \frac{\alpha x}{2} - \frac{\alpha-1}{2} \ln x - \int \mathrm{d}\lambda \, \sigma(\lambda) \ln(x-\lambda) - \frac{1}{2} \Big[1+ \frac{1}{\alpha} + \ln \alpha\Big].
\end{align}
The logarithmic potential of the Marchenko-Pastur law is known analytically, as stated in Proposition~II.1.5 of \cite{faraut2014logarithmic}. 
More precisely, we have for all $x \geq \lambda_+(\alpha)$:
\begin{align*}
    \int & \mathrm{d}\lambda \, \sigma(\lambda) \ln(x-\lambda)\\ 
     &= \frac{\alpha x}{2} - \frac{\alpha-1}{2} \ln x - \frac{1}{2} \Big[1 + \frac{1}{\alpha} + \ln \alpha \Big] - \frac{\alpha}{2} \int_{\lambda_+(\alpha)}^x \frac{\sqrt{(u - \lambda_+(\alpha))(u-\lambda_-(\alpha))}}{u} \ \mathrm{d}u.
\end{align*}
It is then immediate to see that eq.~\eqref{eq:rate_MP_2} and eq.~\eqref{eq:rate_MP} are equivalent, 
validating thus our general result in this case.

\section{\label{sec_app:phase_transition}The phase transition in the rate function}

\noindent
In this section, we investigate possible discontinuities in the derivatives of the rate function $I(x)$, when $d_\mathrm{max} > 0$ and $x_c(\rho)$ is finite. In this case, the function 
$\overline{G}_\sigma(x)$ is constant and equal to $\alpha / d_\mathrm{max}$ for $x \geq x_c(\rho)$.
Recall that if $s_\mathrm{max} \leq x \leq x_c(\rho)$, $\overline{G}_\sigma(x)$ is the second branch to the Marchenko-Pastur equation (eq.~(1) of the main text). This equation can be written as $F_\sigma(G) = x$, with
\begin{align}\label{eq:F_sigma}
  F_\sigma(G) &= \frac{1}{G} + \alpha \int \mathrm{d}t \rho(t) \frac{t}{\alpha - t G}.
\end{align}
Moreover we know that $G_\sigma(s_\mathrm{max}) \leq \overline{G}_\sigma(x) \leq \alpha/d_\mathrm{max}$.
By differentiating the relation $F_\sigma(\overline{G}_\sigma(x)) = x$, we find 
\begin{align*}
\overline{G}_\sigma'(x) = 1/F_\sigma'(\overline{G}_\sigma(x)).
\end{align*}
Let us assume that $\rho(t) \sim (d_\mathrm{max}-t)^\eta$ with $\eta > 0$ and $t$ close to $d_\mathrm{max}$.
If $\eta \geq 1$, we have $G_\rho'(d_\mathrm{max}) < \infty$, so that $F_\sigma'(\alpha/d_\mathrm{max}) < \infty$, and $\overline{G}_\sigma'(x) \to 1/F_\sigma'(\alpha/d_\mathrm{max}) > 0$ as $x \uparrow x_c(\rho)$.
The transition in $I(x)$ is thus of second order in this case, as $\overline{G}_\sigma'(x)$ is discontinuous.
\\
If we now assume that $\eta < 1$, we have $G_\rho'(d_\mathrm{max}) = +\infty$.
By eq.~\eqref{eq:F_sigma}, this implies that $\overline{G}_\sigma'(x) \to 0$ as $x \uparrow x_c(\rho)$.
Thus in this case both $\overline{G}_\sigma$ and $\overline{G}_\sigma'$ are continuous in $x = x_c(\rho)$.
We can differentiate the relation $F_\sigma(\overline{G}_\sigma(x)) = x$ once more, and we find easily:
\begin{align}\label{eq:Gsigma_second}
    \overline{G}_\sigma''(x) &= -\frac{F_\sigma''(\overline{G}_\sigma(x))}{F_\sigma'(\overline{G}_\sigma(x))^3}.
\end{align}
From eqs.~\eqref{eq:F_sigma},~\eqref{eq:Gsigma_second}, one can show that $\overline{G}_\sigma''(x) \to 0$ as $x \to x_c(\rho)$ if and only if $\eta < 1/2$. In particular, for 
any $1/2 \leq \eta < 1$, the transition in $I(x)$ is of third order. As mentioned in the main text, similar transitions and their dependency on the vanishing exponent of the density are discussed in \cite{majumdar2014top}, 
in the context of multi-critical matrix models.
\\
Differentiating three times, one can show in a similar way that the transition is of fourth order if and only if $\eta \in [1/3,1/2)$. 
Generalizing this to any order, we conjecture that $I(x)$ is smooth at any point $x \neq x_c(\rho)$, and 
that the first discontinuous derivative of the rate function at $x = x_c(\rho)$ is $I^{(k+1)}(x)$, with $\eta \in [1/k,1/(k-1))$ (with the convention $1/0 = +\infty$).

\section{Technical details of the derivation}\label{sec_app:technical}

\subsection{The law of \texorpdfstring{$H_n$}{Hn} under the first tilt}\label{subsec_app:tilt_Hn}

\noindent Recall the tilted distribution (with $\mathcal{D}\bz \equiv \mathrm{d}\bz \ e^{-||\bz||^2/2} / (2 \pi)^{n/2}$ the standard Gaussian law):
\begin{align}
   P_{\theta,\bbe}(\bz) \mathrm{d}\bz \propto \mathcal{D}\bz \ e^{\frac{\theta n}{2} \bbe^\intercal H_n \bbe}.
\end{align}
Computing the normalization factor, we reach that:
\begin{align}\label{eq_app:tilted_distribution}
  P_{\theta,\bbe}(\bz) &= \exp\Big\{ \frac{1}{2} \sum_{\mu=1}^m \ln \Big(1 - \frac{\theta}{\alpha} d_\mu\Big) + \frac{ \theta}{2 \alpha} \sum_{\mu=1}^m d_\mu (\bbe^\intercal \bz_\mu)^2 - \frac{1}{2} \sum_{\mu=1}^m \norm{\bz_\mu}^2 - \frac{nm}{2} \ln 2 \pi\Big\}, \nonumber \\
  &= \prod_{\mu=1}^m \exp\Big\{- \frac{1}{2} \bz_\mu^\intercal(\mathds{1}_n - \frac{\theta}{\alpha} d_\mu \bbe \bbe^\intercal) \bz_\mu - \frac{n}{2} \ln 2 \pi + \frac{1}{2} \ln \det \big(\mathds{1}_n - \frac{\theta}{\alpha} d_\mu \bbe \bbe^\intercal\big)\Big\}.
\end{align}
The matrix $\mathds{1}_n - \frac{\theta}{\alpha} d_\mu \bbe \bbe^\intercal$ is a rank-one modification of the identity, so we easily compute
\begin{align}\label{eq_app:1}
  \big(\mathds{1}_n - \frac{\theta}{\alpha} d_\mu \bbe \bbe^\intercal\big)^{-1/2} &= \mathds{1}_n + \big((1 - \theta d_\mu / \alpha)^{-1/2} - 1\big) \bbe \bbe^\intercal.
\end{align}
Changing variables $\bz'_\mu = \big(\mathds{1}_n - \theta d_\mu \bbe \bbe^\intercal / \alpha\big)^{1/2} \bz_\mu$ in eq.~\eqref{eq_app:tilted_distribution} and using eq.~\eqref{eq_app:1} yields the law of $H_n^{(\bbe,\theta)}$ in the main text.

\subsection{Simplifying \texorpdfstring{$J_1(\theta,x)$}{J1}}\label{subsec_app:simplifiying_J1}

In this section, we simplify the expression of $J_1(\theta,x)$ when $\theta \leq \theta_c(x) \equiv G_\sigma(x)$. We start from eq.~(7) of the main text, that we recall here:
\begin{equation}\label{eq:J1_variational_app}
   J_1(\theta,x) = \inf_{\gamma > \theta x} \big[\frac{\gamma}{2} - \frac{1}{2} \int \mathrm{d}u \, \sigma(u) \ln (\gamma - \theta u)\big] - \frac{1}{2}.
\end{equation}
As we saw in the main text, when $\theta \leq G_\sigma(x)$ the infimum is reached in $\gamma^\star = \theta G_\sigma^{-1}(\theta)$.
This implies
\begin{equation}
   J_1(\theta,x) = \frac{\theta G_\sigma^{-1}(\theta)}{2} - \frac{1}{2} \ln \theta - \frac{1}{2} \int \mathrm{d}u \, \sigma(u) \ln (G_\sigma^{-1}(\theta) - u)\big] - \frac{1}{2}.
\end{equation}
Let us differentiate this expression with respect to $\theta$:
\begin{align*}
   \partial_\theta J_1(\theta,x) &= \frac{G_\sigma^{-1}(\theta)}{2} + \frac{\theta}{2 G_\sigma'(G_\sigma^{-1}(\theta))} - \frac{1}{2 \theta} - \frac{G_\sigma(G_\sigma^{-1}(\theta))}{2 G_\sigma'(G_\sigma^{-1}(\theta))}, \\
   &= \frac{G_\sigma^{-1}(\theta)}{2} - \frac{1}{2 \theta}.
\end{align*}
Using now the Marchenko-Pastur equation (eq.~(1) of the main text), we can simplify this into:
\begin{align*}
   \partial_\theta J_1(\theta,x) &= \alpha \int \mathrm{d}t \, \rho(t) \, \frac{t}{\alpha - t \theta} = F_1'(\theta). 
\end{align*}
Since $J_1(0,x) = F_1(0) = 0$, this implies that for every $\theta \leq \theta_c(x)$ we have $J_1(\theta,x) = F_1(\theta)$, which justifies the claim in the main text.

\subsection{Derivations of \texorpdfstring{$F_2(\theta)$}{F2} and \texorpdfstring{$J_2(\theta,x)$}{J2}}\label{subsec_app:derivation_transition_F2_J2}

\subsubsection{The derivation of \texorpdfstring{$F_2(\theta)$}{F2}}

We start from the definition of $F_2(\theta)$ (omitting the limit $n \to \infty$, that we will take in the end):
\begin{align*}
    F_2(\theta) = \frac{1}{n} \ln \int \mathcal{D}\bz \int_{\norm{\bbe}^2 = 1} \hspace{-0.4cm} \mathrm{d}\bbe \int_{\norm{\bbf}^2 = 1} \hspace{-0.4cm}\mathrm{d}\bbf  \ e^{\frac{\theta n}{\sqrt{m}} \sum\limits_{i,\mu} \sqrt{d_\mu} e_i z_{\mu i} f_\mu}.
\end{align*}
Integrating over $\bz$ yields:
\begin{align*}
    F_2(\theta) &= \frac{1}{n} \ln \int_{\norm{\bbf}^2 = 1} \hspace{-0.4cm}\mathrm{d}\bbf  \ e^{\frac{\theta^2 n^2}{2m} \sum\limits_{\mu} d_\mu f_\mu^2}, \\
    &= \frac{1}{n} \ln \frac{\int \mathrm{d}\bbf \, \delta(||\bbf||^2 - m)  \ e^{\frac{\theta^2 n^2}{2m^2} \sum\limits_{\mu} d_\mu f_\mu^2}}{\int \mathrm{d}\bbf \, \delta(||\bbf||^2 - m)},
\end{align*}
in which we rescaled the norm of $\bbf$.
We introduce a Lagrange multiplier $\gamma$ to fix the norm of $\bbf$. This yields (recall $\alpha = m/n$):
\begin{align*}
   F_2(\theta) &= \inf_\gamma \Big[\frac{\alpha \gamma}{2} - \frac{1}{2n} \ln \det (\gamma \mathds{1}_m - \frac{\theta^2}{\alpha^2} D_m)\Big] - \frac{\alpha}{2}.
\end{align*}
Since the matrix inside the log-det must be positive, we have $\gamma \geq \theta^2 d_\mathrm{max} / \alpha^2$.
Changing variable, by letting $\gamma = \theta^2 \gamma' / \alpha^2$, we arrive at, when $n \to \infty$:
\begin{align*}
   F_2(\theta) &= \frac{\alpha}{2} \inf_{\gamma \geq d_\mathrm{max}} \Big[\frac{\theta^2 \gamma}{\alpha^2} - \int \mathrm{d}t \, \rho(t) \, \ln(\gamma - t)\Big] - \frac{\alpha}{2} \ln \frac{\theta^2}{\alpha^2} - \frac{\alpha}{2}.
\end{align*}
This ends the derivation of the expression of $F_2(\theta)$ given in the main text.

\subsubsection{Computing \texorpdfstring{$J_2(\theta,x)$}{J2}}
The goal of this section is to compute $J_2(\theta,x)$. More precisely, we will show eq.~\eqref{eq:J2_Lagrange}, which will then be simplified in the following section, 
precisely showing the transition phenomenon described in the main text. 
We start from the definition of $J_2(\theta,x)$ (we omit the $n \to \infty$ limit for simplicity, we will take it at the end):
\begin{align*}
   J_2(\theta,x) &= \frac{1}{n} \ln \int_{\norm{\bbe}^2 = 1} \hspace{-0.4cm} \mathrm{d}\bbe \int_{\norm{\bbf}^2 = 1} \hspace{-0.4cm}\mathrm{d}\bbf  \ e^{\frac{\theta n}{\sqrt{m}} \sum\limits_{i,\mu} \sqrt{d_\mu} e_i z_{\mu i} f_\mu}, \\
   &= \frac{1}{n} \ln \frac{\int \mathrm{d}\bbe \int\mathrm{d}\bbf \ \delta(||\bbe||^2-n) \ \delta(||\bbf^2|| - m)  \ e^{\theta \frac{\sqrt{n}}{m} \sum\limits_{i,\mu} \sqrt{d_\mu} e_i z_{\mu i} f_\mu}}{\int \mathrm{d}\bbe \int\mathrm{d}\bbf \ \delta(||\bbe||^2-n) \ \delta(||\bbf^2|| - m)}.
\end{align*}
We introduce two Lagrange multipliers to fix the norms of $\bbe$ and $\bbf$. Let us start with the computation of the denominator:
\begin{align}
   \frac{1}{n} \ln \int \mathrm{d}\bbe \int\mathrm{d}\bbf \ \delta(||\bbe||^2-n) \ \delta(||\bbf^2|| - m) &\simeq \inf_{\Lambda_e,\Lambda_f \geq 0}\Big[\frac{\Lambda_e}{2} + \frac{\alpha \Lambda_f}{2} - \frac{1}{2} \ln \Lambda_e - \frac{\alpha}{2} \ln \Lambda_f + \frac{(1+\alpha)}{2} \ln 2 \pi\Big].
\end{align}
The positivity constraint on $\Lambda_e,\Lambda_f$ arises naturally for the Gaussian integral to be well-defined. This is easily solved by $\Lambda_e = \Lambda_f = 1$, and we arrive at:
\begin{align}\label{eq:denominator}
   \frac{1}{n} \ln \int \mathrm{d}\bbe \int\mathrm{d}\bbf \ \delta(||\bbe||^2-n) \ \delta(||\bbf^2|| - m) &\simeq \frac{(1+\alpha)}{2} (1 + \ln 2 \pi).
\end{align}
We use the same method to compute the numerator:
\begin{align}
   &\frac{1}{n} \ln \int \mathrm{d}\bbe \int\mathrm{d}\bbf \ \delta(||\bbe||^2-n) \ \delta(||\bbf^2|| - m)  \ e^{\theta \frac{\sqrt{n}}{m} \sum\limits_{i,\mu} \sqrt{d_\mu} e_i z_{\mu i} f_\mu} \\ 
   &\simeq \inf_{\Lambda_e,\Lambda_f \geq 0}\Big[\frac{\Lambda_e}{2} + \frac{\alpha \Lambda_f}{2} - \frac{1}{2n} \ln \det \begin{pmatrix}
      \Lambda_e \mathds{1}_n & \frac{\theta}{\sqrt{\alpha}} \frac{\bz^\intercal}{\sqrt{m}} \sqrt{D_m}  \\ 
      \frac{\theta}{\sqrt{\alpha}} \sqrt{D_m} \frac{\bz}{\sqrt{m}} & \Lambda_f \mathds{1}_m 
   \end{pmatrix} + \frac{(1+\alpha)}{2} \ln 2 \pi\Big] \nonumber.
\end{align}
We can compute the determinant of the block matrix easily, and we arrive at:
\begin{align}
   &\frac{1}{n} \ln \int \mathrm{d}\bbe \int\mathrm{d}\bbf \ \delta(||\bbe||^2-n) \ \delta(||\bbf^2|| - m)  \ e^{\theta \frac{\sqrt{n}}{m} \sum\limits_{i,\mu} \sqrt{d_\mu} e_i z_{\mu i} f_\mu} \\ 
   &\simeq \inf_{\Lambda_e,\Lambda_f \geq 0}\Big[\frac{\Lambda_e}{2} + \frac{\alpha \Lambda_f}{2} -  \frac{\alpha -1 }{2} \ln \Lambda_f - \frac{1}{2n} \ln \det \big(\Lambda_e \Lambda_f \mathds{1}_n - \frac{\theta^2}{\alpha} H_n\big) + \frac{(1+\alpha)}{2} \ln 2 \pi\Big] \nonumber.
\end{align}
Note that the matrix inside the log-det must be positive, which constrains $\Lambda_e \Lambda_f \geq \theta^2 x / \alpha$, as we assumed that the largest eigenvalue of $H_n$ converges to $x$ as 
$n \to \infty$. All in all, we have, taking $n \to \infty$:
\begin{align}\label{eq:numerator}
   &\frac{1}{n} \ln \int \mathrm{d}\bbe \int\mathrm{d}\bbf \ \delta(||\bbe||^2-n) \ \delta(||\bbf^2|| - m)  \ e^{\theta \frac{\sqrt{n}}{m} \sum\limits_{i,\mu} \sqrt{d_\mu} e_i z_{\mu i} f_\mu} \\ 
   &\simeq \inf_{\substack{\Lambda_e,\Lambda_f \geq 0 \\ \mathrm{s.t.}\ \alpha \Lambda_e \Lambda_f \geq \theta^2 x}}\Big[\frac{\Lambda_e}{2} + \frac{\alpha \Lambda_f}{2} -  \frac{\alpha -1 }{2} \ln \Lambda_f - \frac{1}{2} \int \mathrm{d}\lambda \, \sigma(\lambda) \, \ln \big(\Lambda_e \Lambda_f - \frac{\theta^2}{\alpha} \lambda \big) + \frac{(1+\alpha)}{2} \ln 2 \pi\Big] \nonumber.
\end{align}
Combining eq.~\eqref{eq:numerator} and eq.~\eqref{eq:denominator} yields the sought result:
\begin{align}\label{eq:J2_Lagrange}
   &J_2(\theta,x) \\ 
   &= \frac{1}{2}\inf_{\substack{\Lambda_e,\Lambda_f \geq 0 \\ \mathrm{s.t.}\ \alpha \Lambda_e \Lambda_f \geq \theta^2 x}}\Big[\Lambda_e + \alpha \Lambda_f - (\alpha - 1) \ln \Lambda_f - \int \mathrm{d}\lambda \, \sigma(\lambda) \ln \big(\Lambda_e\Lambda_f - \frac{\theta^2}{\alpha} \lambda\big) \Big] - \frac{1+\alpha}{2}.\nonumber
\end{align}

\subsubsection{The transition in \texorpdfstring{$J_2(\theta,x)$}{J2}}
We start from the expression of $J_2(\theta,x)$ of eq.~\eqref{eq:J2_Lagrange}:
\begin{align}\label{eq:J2_1}
   &J_2(\theta,x) \\ 
   &= \frac{1}{2}\inf_{\substack{\Lambda_e,\Lambda_f \geq 0 \\ \mathrm{s.t.}\ \alpha \Lambda_e \Lambda_f \geq \theta^2 x}}\Big[\Lambda_e + \alpha \Lambda_f - (\alpha - 1) \ln \Lambda_f - \int \mathrm{d}\lambda \, \sigma(\lambda) \ln \big(\Lambda_e\Lambda_f - \frac{\theta^2}{\alpha} \lambda\big) \Big] - \frac{1+\alpha}{2}.\nonumber
\end{align}
The variational parameters $\Lambda_e,\Lambda_f$ can saturate, which is associated to a phase transition. At this point $J_2(\theta,x)$ will become sensitive to the largest eigenvalue of $H_n$ (assumed to be equal to $x$).
This phase transition occurs for $\theta = \theta_c(x)$ such that the corresponding values of $\Lambda_e,\Lambda_f$ satisfy $\alpha \Lambda_e \Lambda_f = (\theta_c(x))^2 x$. 
From this equation and the zero-gradient equations on $\Lambda_e,\Lambda_f$ (valid for $\theta\!\leq\!\theta_c(x)$), it is easy to obtain $\theta_c(x) = \sqrt{x G_\sigma(x)^2 + (\alpha-1) G_\sigma(x)}$.

{\it The case $\theta \leq \theta_c(x)$ - } In this case $J_2(\theta,x)$ is not sensitive to the value of $x$, and we can use a very useful expression derived in \cite{maillard2020landscape} for the log-potential of $\sigma(\lambda)$. For any $x \geq s_\mathrm{max}$:
\begin{align*}
   \int \mathrm{d}\lambda \, \sigma(\lambda) \ln(x-\lambda) &= \inf_{0 < g < G_\sigma(s_\mathrm{max})} \Big[- \ln g + z g + \alpha \int \mathrm{d}t \, \rho(t) \ln(\alpha - t g)\Big] - 1 - \alpha \ln \alpha.
\end{align*}
Note that this infimum is attained at $g = G_\sigma(x)$, as this value is the unique zero of the derivative of the expression above, by eq.~\eqref{eq:F_sigma}.
We can write from eq.~\eqref{eq:J2_1}:
\begin{align}
    \label{eq:J2_first_phase_1}
   J_2(\theta,x) &=  -  \frac{\alpha(1- \ln \alpha)}{2} - \frac{1}{2} \ln \frac{\theta^2}{\alpha}+ \frac{1}{2}\inf_{0 < g < G_\sigma(s_\mathrm{max})}\inf_{\substack{\Lambda_e,\Lambda_f \geq 0 \\  (\alpha \Lambda_e \Lambda_f \geq \theta^2 x)}} \Big[\Lambda_e + \alpha \Lambda_f \\
   &- (\alpha - 1) \ln \Lambda_f + \ln g - \frac{\alpha \Lambda_e \Lambda_f}{\theta^2} g - \alpha \int \mathrm{d}t \, \rho(t) \ln(\alpha - t g) \Big] \nonumber.
\end{align}
Since we are in the ``no-saturation'' regime, we can use the zero-gradient equations on $\Lambda_e,\Lambda_f$:
\begin{align*}
    \begin{cases}
    \Lambda_f &= \theta^2 / (\alpha g), \\ 
    \Lambda_e &= \theta^2/ g - (\alpha-1).
    \end{cases}
\end{align*}
Plugging this back into eq.~\eqref{eq:J2_first_phase_1} we obtain:
\begin{align*}
   J_2(\theta,x) &= \frac{1}{2} \inf_{0 < g < G_\sigma(s_\mathrm{max})}\Big[\frac{\theta^2}{g} + \alpha \ln g - \alpha \ln \frac{\theta^2}{\alpha}  - \alpha \int \mathrm{d}t \, \rho(t) \ln(\alpha - t g) + \alpha (\ln \alpha - 1)\Big].
\end{align*}
Changing variables $\gamma = \alpha/g$, we reach:
\begin{align}\label{eq:J2_first_phase_2}
   J_2(\theta,x) &= \frac{\alpha}{2}\inf_{\gamma \geq \alpha / G_\sigma(s_\mathrm{max})}\Big[\frac{\gamma \theta^2}{\alpha^2} - \int \mathrm{d}t \rho(t) \ln (\gamma - t)\Big] - \frac{\alpha}{2} \Big(1+ \ln \frac{\theta^2}{\alpha^2}\Big).
\end{align}
In order to map $J_2(\theta,x)$ to $F_2(\theta)$, we must only show that the Lagrange multiplier $\gamma$  in eq.~\eqref{eq:J2_first_phase_2} does not ``saturate'' 
for $\theta \leq \theta_c(x)$.
This is easily shown using eq.~\eqref{eq:F_sigma}.
Since $\theta \leq \theta_c(x)$, we also have $\theta \leq \theta_c(s_\mathrm{max})$, and thus:
\begin{align*}
   \theta^2 &\leq s_\mathrm{max} G_\sigma(s_\mathrm{max})^2 + (\alpha-1)G_\sigma(s_\mathrm{max}) \leq \alpha G_\sigma(s_\mathrm{max})\Big[1+G_\sigma(s_\mathrm{max})\Big(\frac{s_\mathrm{max}}{\alpha} - \frac{1}{\alpha G_\sigma(s_\mathrm{max})}\Big) \Big], \\ 
   &\leq \alpha G_\sigma(s_\mathrm{max})\Big[1+G_\sigma(s_\mathrm{max}) \int \frac{\mathrm{d}t \, \rho(t) \, t}{\alpha - t G_\sigma(s_\mathrm{max})} \Big] \leq \alpha^2 \int \frac{\mathrm{d}t \, \rho(t)}{\alpha/G_\sigma(s_\mathrm{max})- t}.
\end{align*}
This precisely means that the infimum in eq.~\eqref{eq:J2_first_phase_2} will be attained for a point $\gamma$ which is a critical point of the functional inside the infimum: 
\begin{align*}
   \frac{\theta^2}{\alpha^2} = \int \mathrm{d}t \, \rho(t) \frac{1}{\gamma-t},
\end{align*}
i.e.\ there is no saturation, and we have $J_2(\theta,x) = F_2(\theta)$.

{\it The case $\theta \geq \theta_c(x)$ - } 
In this case, we have a ``saturation'' in the infimum of eq.~\eqref{eq:J2_1}.
More precisely, the $\Lambda_e,\Lambda_f$ attaining the infimum satisfy $\alpha \Lambda_e \Lambda_f = \theta^2 x$.
One can solve the infimum over $\Lambda_e,\Lambda_f$ constrained by this equality. 
Introducing a Lagrange parameter $\rho$, we reach:
\begin{align*}
   J_2(\theta,x) &=- \frac{1+\alpha}{2} - \frac{1}{2} \ln \frac{\theta^2}{\alpha} - \frac{1}{2} \int \mathrm{d}\lambda \, \sigma(\lambda) \ln (x - \lambda) \\
   &+\frac{1}{2}\inf_{\Lambda_e,\Lambda_f \geq 0} \extr_\rho \Big[\Lambda_e + \alpha \Lambda_f - (\alpha-1)\ln \Lambda_f - \rho \Big(\Lambda_e \Lambda_f - \frac{\theta^2 x}{\alpha}\Big)\Big].
\end{align*}
The $\extr$ notation denotes solving the associated zero-gradient equation, as is standard with Lagrange multipliers. One can now solve the infimum over $\Lambda_e,\Lambda_f$ easily, and we reach:
\begin{align}\label{eq:J2_second_phase_1}
   J_2(\theta,x) &=\frac{1}{2} \extr_\rho \Big[\frac{\alpha}{\rho} + (\alpha-1) \ln \rho + \frac{ \rho \theta^2 x}{\alpha} - \ln \frac{\theta^2}{\alpha} - \int \mathrm{d}\lambda \, \sigma(\lambda) \ln (x - \lambda) \Big] - \frac{1+\alpha}{2}.
\end{align}
This can also be solved easily, and finally we have, for $\theta \geq \theta_c(x)$:
\begin{align}\label{eq:J2_second_phase_2}
   J_2(\theta,x) &=\frac{1}{2}\Big[- (1+\alpha) - \alpha \ln \frac{\theta^2}{\alpha} - (\alpha-1) \ln (2x) + \sqrt{(\alpha-1)^2+4x\theta^2}  \nonumber \\
   & \hspace{1.0cm} + (\alpha-1) \ln \big[1-\alpha + \sqrt{(\alpha-1)^2+4x\theta^2}\big] - \int \mathrm{d}\lambda \, \sigma(\lambda) \ln (x - \lambda) \Big].
\end{align}
This ends the argument by justifying all the expressions given for $J_2(\theta,x)$ in the main text.

{\it The case $d_\mathrm{max} \leq 0$ - } In this case (not considered in the calculation of $J_2$) $s_\mathrm{max} \leq 0$, and 
the transition we described does not take place, as $\Lambda_e,\Lambda_f \geq 0$ can not satisfy $\alpha \Lambda_e \Lambda_f = \theta^2 x < 0$. 
The difference between the quenched and annealed integrals in this case has, as far as we know, not been investigated before, and it remains an open question. 
In this setting the first ``naive'' tilting allows to derive the large deviations, as emphasized in the main text.

\section{Simplifying the rate function \texorpdfstring{$I(x)$}{I(x)}}\label{sec_app:simplify_rate}

\subsection{When \texorpdfstring{$x < x_\mathrm{max}$}{x<xmax}}\label{subsec_app:simplify_rate_dmax_negative}

\noindent
The goal of this section is to show, for all $x < x_\mathrm{max}$:
\begin{align}
   \underbrace{\sup_{\theta} [J_1(\theta,x) - F_1(\theta)]}_{I(x)} & = \frac{1}{2} \int_{s_\mathrm{max}}^x [\overline{G}_\sigma(u) - G_\sigma(u)] \mathrm{d}u,
\end{align}
and that the maximum is reached in $\theta_x = \overline{G}_\sigma(x)$.
Recall eq.~(15) of the main text:
\begin{align}
   J_1(\theta,x)= 
   \begin{cases}
        F_1(\theta) = -\frac{\alpha}{2} \int \mathrm{d}t \rho(t) \ln (1 - \alpha^{-1} \theta t) &\mathrm{if} \ \theta \leq G_\sigma(x), \\    
        \frac{\theta x - 1 - \ln \theta}{2} - \frac{1}{2}\int \mathrm{d}\lambda \ \sigma(\lambda) \ln(x - \lambda) &\mathrm{if} \ \theta \geq G_\sigma(x).
   \end{cases} 
\end{align}
Differentiating with respect to $\theta$ in this equation, we reach:
\begin{align}
   \partial_\theta[J_1(\theta,x) - F_1(\theta)] &= 
   \begin{cases}
    0 &\mathrm{if} \ \theta \leq G_\sigma(x), \\ 
    \frac{\theta x-1}{2 \theta} - \frac{\alpha}{2} \int \mathrm{d}t \, \rho(t) \frac{t}{\alpha - \theta t} &\mathrm{if} \ \theta > G_\sigma(x).
   \end{cases} 
\end{align}
So the supremum $\sup_{\theta \geq 0} [J_1(\theta,x) - F_1(\theta)]$ is attained for $\theta = \theta_x > G_\sigma(x)$ that satisfies:
\begin{align*}
    x &= \frac{1}{\theta} + \alpha \int \mathrm{d}t \rho(t) \frac{t}{\alpha - \theta t}.
\end{align*}
Note that this is exactly the Marchenko-Pastur equation~\eqref{eq:F_sigma}, 
so that $\theta_x$ is the second ``branch'' to the Marchenko-Pastur equation, i.e.\ precisely $\theta_x = \overline{G}_\sigma(x)$.
Moreover, we know that $I(s_\mathrm{max}) = 0$, and we conclude by noticing that:
\begin{align*}
    I'(x) &= \partial_x[J_1(\theta_x,x) - F_1(\theta_x)] = \frac{\theta_x}{2} - \frac{1}{2} \int \frac{\mathrm{d}\lambda \, \sigma(\lambda)}{x - \lambda} = \frac{1}{2}[\overline{G}_\sigma(x) - G_\sigma(x)].
\end{align*}

\subsection{The second tilting}\label{subsec_app:simplify_rate_dmax_positive}

\noindent
Our objective is to show, for all $x \geq s_\mathrm{max}$:
\begin{align}\label{eq:to_show_2}
   I(x) = \sup_{\theta \geq 0} [J_2(\theta,x) - F_2(\theta)] &= \frac{1}{2} \int_{s_\mathrm{max}}^x [\overline{G}_\sigma(u) - G_\sigma(u)] \mathrm{d}u. 
\end{align}
Recall the functions $J_2$ and $F_2$ (with $\theta_c(x) = \sqrt{x G_\sigma(x)^2 + (\alpha-1)G_\sigma(x)}$):
\begin{align}
   J_2(\theta,x)= 
   \begin{cases}
        F_2(\theta) = \frac{\alpha}{2} \inf_{\gamma \geq d_\mathrm{max}} \hspace{-0.1cm} \Big[\frac{\gamma \theta^2}{\alpha^2} - \int \mathrm{d}t\rho(t) \ln(\gamma-t) - 1 - \ln\frac{\theta^2}{\alpha^2}\Big] &\mathrm{if} \ \theta \leq \theta_c(x), \\    
        \frac{\alpha-1}{2} \ln \big[\frac{1-\alpha + \sqrt{(\alpha-1)^2+4x\theta^2}}{2x}\big]\!-\!\frac{1+\alpha}{2} -\!\frac{\alpha}{2} \ln \frac{\theta^2}{\alpha}\! &\\ 
        \hspace{2cm} +\!\frac{1}{2}\sqrt{(\alpha-1)^2+4x\theta^2} - \frac{1}{2}\!\int\!\mathrm{d}\lambda \,\sigma(\lambda) \ln (x - \lambda) &\mathrm{if} \ \theta \geq \theta_c(x).
   \end{cases} 
\end{align}
We perform the change of variable $\theta(\tau,x)^2 \equiv x \tau^2 + (\alpha-1)\tau$. 
At the critical value $\theta_c(x)$, we have $\tau_c(x) = G_\sigma(x)$.
We obtain the expression of the rate function as $I(x) = \sup_{\tau \geq G_\sigma(x)} I(x,\tau)$, with
$I(x,\tau) = J(\tau,x) - F(\tau,x)$, in which we 
naturally defined:
\begin{align}\label{eq:def_J_tau}
   J(\tau,x) = &= \frac{1}{2}\Big\{\!- 2-\!\alpha\!\ln \Big[\frac{\tau x}{\alpha} + 1 - \frac{1}{\alpha}\Big]- \ln (\tau) + 2 x \tau - \! \int \mathrm{d}\lambda \, \sigma(\lambda) \ln (x - \lambda) \Big\}.
\end{align}
Note that we have the following expression for $F(\tau,x)$:
\begin{align}\label{eq:def_F_tau}
    F(\tau,x) &= \\
    & \nonumber 
    \begin{cases}
        \frac{\alpha}{2} \int_0^{\theta(\tau,x)^2/\alpha^2} \big[G_\rho^{-1}(u) - \frac{1}{u}\big] \mathrm{d}u&\mathrm{if} \ \theta(\tau,x)^2 \leq \alpha^2 G_\rho(d_\mathrm{max}), \\
        \frac{\alpha}{2} [\frac{d_\mathrm{max} \theta(\tau,x)^2}{\alpha^2} - \int \mathrm{d}t \rho(t) \ln (d_\mathrm{max} - t) - 1 - \ln \frac{\theta(\tau,x)^2}{\alpha^2}] &\mathrm{if} \ \theta(\tau,x)^2 \geq \alpha^2 G_\rho(d_\mathrm{max}).
    \end{cases}
\end{align}
We then compute $\tau_x \equiv \mathrm{argmax}_{\tau \geq G_\sigma(x)} [J(\tau,x) - F(\tau,x)]$ using the derivatives of eqs.~\eqref{eq:def_J_tau},\eqref{eq:def_F_tau}:
\begin{align}\label{eq:dJFtau} 
    \partial_\tau &[J(\tau,x) - F(\tau,x)] = \\ 
    & \nonumber\begin{cases}
        \frac{(2 \tau x + \alpha -1)}{2 \alpha  \tau }(\alpha -\tau G_\rho^{-1}[\theta(\tau,x)^2/\alpha^2]) &\mathrm{if} \ \theta_c(x)^2 \leq\!\theta(\tau,x)^2\!\leq\!\alpha^2 G_\rho(d_\mathrm{max}), \\
        \frac{(2 \tau x + \alpha -1)}{2 \alpha  \tau }\left(\alpha -\tau d_\mathrm{max}\right) &\mathrm{if} \ \theta(\tau,x)^2 \geq \alpha^2 G_\rho(d_\mathrm{max}).
    \end{cases}
\end{align}
For all $s_\mathrm{max} \leq x \leq x_c(\rho) \equiv d_\mathrm{max} G_\rho(d_\mathrm{max})^2 + (\alpha^{-1}-1)d_\mathrm{max}$, the equation $\alpha\!=\!\tau G_\rho^{-1}[\theta(\tau,x)^2/\alpha^2]$ is again the Marchenko-Pastur equation~\eqref{eq:F_sigma}, with $G = \tau$.
Since $\tau_x > G_\sigma(x)$, it is easy to check from eq.~\eqref{eq:dJFtau} that the supremum is attained in $\tau_x = \overline{G}_\sigma(x)$. 
This is true even if $x > x_c(\rho)$, as the maximum is attained in $\tau = \alpha/d_\mathrm{max} = \overline{G}_\sigma(x)$, again from eq.~\eqref{eq:dJFtau}.
Moreover we can compute from eq.~\eqref{eq:to_show_2}:
\begin{align*}
  I'(x) &= \partial_x (J - F)(\tau_x,x) + (\partial_x \tau_x) \underbrace{\partial_\tau (J - F)(\tau_x,x)}_{=0} = \partial_x [J-F](\tau_x,x) = \frac{1}{2} [\tau_x - G_\sigma(x)].
\end{align*}
Since $I(s_\mathrm{max}) = 0$ and $\tau_x = \overline{G}_\sigma(x)$, this last equality implies eq.~\eqref{eq:to_show_2}.

\end{document}